\documentclass[useAMS,usenatbib]{mn2e}
\usepackage{epsfig}
\usepackage{amsmath}

\newcommand{\be}{\begin{equation}}
\newcommand{\beq}{\begin{equation}}
\newcommand{\ba}{\begin{eqnarray}}
\newcommand{\ee}{\end{equation}}
\newcommand{\eeq}{\end{equation}}
\newcommand{\ea}{\end{eqnarray}}

\def\lsim{~\rlap{$<$}{\lower 1.0ex\hbox{$\sim$}}}

\def\gsim{~\rlap{$>$}{\lower 1.0ex\hbox{$\sim$}}}

\defcitealias{geil07}{GW07}

\title[Quasar HII Regions]{The effect of Galactic foreground subtraction on redshifted 21-cm observations of quasar HII regions}

\author[Geil et al.]{Paul. M. Geil$^1$\thanks{Email: pgeil@physics.unimelb.edu.au}, J. Stuart B. Wyithe$^1$, Nada Petrovic$^2$, S. Peng Oh$^2$ \\
$^1$School of Physics, University of Melbourne, Parkville, Victoria,
Australia\\
$^2$Department of Physics, University of California,Santa Barbara, CA 93106, USA.}

\begin{document}

%\date{\today}
%\pagerange{\pageref{firstpage}--\pageref{lastpage}} \pubyear{2006}

\maketitle

\label{firstpage}
\begin{abstract}

We assess the impact of Galactic synchrotron foreground removal on the
observation of high-redshift quasar HII regions in redshifted 21-cm
emission. We consider the case where a quasar is observed in
an intergalactic medium (IGM) whose ionisation structure evolves
slowly relative to the light crossing time of the HII region, as well
as the case where the evolution is rapid. The latter case is
expected towards the end of the reionisation era where the highest
redshift luminous quasars will be observed. In the absence of
foregrounds the fraction of neutral hydrogen in the IGM could be
measured directly from the contrast between the HII region and
surrounding IGM. However, we find that foreground removal lowers the
observed contrast between the HII region and the IGM. This indicates
that measurement of the neutral fraction would require modelling to
correct for this systematic effect. On the other hand, foreground removal does
not modify the most prominent features of the 21-cm maps. Using a
simple algorithm we demonstrate that measurements of the size and
shape of observed HII regions will not be affected by continuum
foreground removal. Moreover, measurements of these quantities will
not be adversely affected by the presence of a rapidly evolving IGM.

\end{abstract}

\begin{keywords}
cosmology: diffuse radiation, large-scale structure, theory -- galaxies: high redshift, intergalactic medium
\end{keywords}

\section{Introduction}
\label{intro} 

The reionisation of cosmic hydrogen by the first stars and galaxies
was an important milestone in the history of the Universe
(e.g. \citeauthor{bl2001} \citeyear{bl2001}).  A powerful tool for
study of the reionisation history will be provided by the redshifted
21-cm emission from neutral hydrogen in the intergalactic medium
(IGM), and several probes of the reionisation era in redshifted
21-cm emission have been suggested. These include observation of the
emission as a function of redshift averaged over a large area of
sky. This observation would provide a direct probe of the evolution in
the neutral fraction of the IGM, and is referred to as the global step
\citep{shaver1999,gnedin2004,furl2006}. Unless reionisation is
uniform throughout the whole IGM however, the global step will be very
difficult to detect beneath the bright foreground. A more powerful
probe will be provided by observation of the 21-cm power spectrum of
fluctuations together with its evolution with redshift.  This
observation would trace the evolution of neutral gas with redshift as
well as the topology of the reionisation process
(e.g. \citeauthor{tozzi2000} \citeyear{tozzi2000};
\citeauthor{furl2004b} \citeyear{furl2004b}; \citeauthor{loeb2004}
\citeyear{loeb2004}; \citeauthor{wm2007} \citeyear{wm2007};
\citeauthor{iliev2006} \citeyear{iliev2006}). Finally, observation of
individual HII regions will probe quasar physics as well as the
evolution of the neutral gas
\citep{wl2004b,kohler2005,valdes2006}. \cite{kohler2005} have
generated synthetic spectra using cosmological simulations and
conclude that quasar HII regions will provide the most prominent
individual cosmological signals. Work by \cite{datta2007b} has focused
on the detection of ionised bubbles in redshifted 21-cm maps.
Their results suggest it may be possible
to blindly detect spherical HII regions of radius$\gsim$ 22 comoving
Mpc during the epoch of reionisation. More recently, \cite{geil07}
(hereafter \citetalias{geil07}) have studied the impact of a
percolating IGM on the detection of HII regions, and shown that
quasars will leave a detectable imprint until very late in the
reionisation era.

Various experiments are planned to measure 21-cm emission from the
pre-reionisation IGM, including the Low Frequency
Array\footnote{http://www.lofar.org/} (LOFAR) and the Murchison
Widefield Array\footnote{http://www.haystack.mit.edu/ast/arrays/mwa/}
(MWA). In addition to their physical configuration, the sensitivity of
these telescopes to the epoch of reionisation will be limited by the
difficulty of achieving a perfect calibration and by problems
associated with foregrounds and the ionosphere. There has been
significant discussion in the literature regarding low-frequency
foregrounds. While fluctuations due to foregrounds are expected to be
orders of magnitude larger than the reionisation signal
(e.g. \citeauthor{dimatteo2002} \citeyear{dimatteo2002};
\citeauthor{oh2003} \citeyear{oh2003}), foreground spectra are
anticipated to be smooth. Since the reionisation signature includes
fluctuations in frequency as well as angle, it has therefore been proposed that
they be removed through continuum subtraction
(e.g. \citeauthor{gnedin2004} \citeyear{gnedin2004};
\citeauthor{wang2006} \citeyear{wang2006}), or using the differences
in symmetry from power spectra analysis
\citep{morales2004,zaldarriaga2004}.

In this paper we focus on the observation of individual HII regions
around known quasars, with emphasis on the capabilities of the MWA. In
particular we address the removal of a bright Galactic foreground from
the redshifted 21-cm signal. Our analysis extends the model and ideas
presented in \citetalias{geil07} by including evolution of the
percolating IGM.  We begin by summarising our semi-numerical model for
the ionisation structure of the IGM (\S\,\ref{model}). We show
examples of model HII regions in \S\,\ref{results} (both with and
without evolution of the IGM). We then discuss the influence of
foregrounds in \S\,\ref{foreground} and describe the observation of
the HII region with a low-frequency array in \S\,\ref{obs}, including
discussion of foreground removal and reconstruction of the HII region
shape. The effects of different foreground removal models on HII
region observables are investigated in \S\,\ref{fg}. We present our conclusions in \S\,\ref{conclusion}.
Throughout we adopt the set of cosmological parameters determined by
WMAP (Spergel et al. \citeyear{spergel2007}) for a flat $\Lambda$CDM universe.

\section{Semi-Numerical Model for the Growth of Quasar HII Regions in a Percolating IGM}
\label{model}

In this section we describe our model for the reionisation of a
three-dimensional volume of the IGM using a semi-numerical scheme
\citep{geil07,zahn2007,mesinger2007}. We follow the implementation
described in \citetalias{geil07}, and refer the reader to that paper
for details of the model. The model employs a semi-analytic
prescription for the reionisation process, which is combined with a
realisation of the density field. Each of these aspects is summarised
briefly below.

\subsection{Density-dependent model of global reionisation}

The semi-analytic component of our model computes the relation between
the local dark matter overdensity and the reionisation of the IGM, and
is based on the model described by \cite{wl2007} and
\cite{wm2007}. The model includes the overabundance of galaxies in
overdense regions that results from galaxy bias \citep{mo1996}, as
well as the increase in the recombination rate in overdense regions,
and ionisation feedback which suppresses low-mass galaxy formation in
ionised regions. This model predicts the sum of astrophysical effects
to be dominated by galaxy bias, and that as a result, overdense
regions are reionised first. This leads to the growth of HII regions
via a phase of percolation during which individual HII regions overlap
around clustered sources in overdense regions of the universe. The
output of the semi-analytic model is evolution of the ionisation
fraction by mass $Q_{\delta,R}$ of a particular region of scale $R$
with overdensity $\delta$.

Throughout this paper we consider a model in which the mean IGM is
reionised at $z=6$ \citep{fan2006,gnedin2006,white2003}. We assume that star
formation proceeds in halos above the hydrogen cooling threshold in
neutral regions of IGM. In ionised regions of the IGM star formation
is assumed to be suppressed by radiative feedback.

\subsection{The ionisation field}
\label{ionisationfield}

We construct an ionisation field based on a Gaussian random field for
the matter overdensity, combined with the value of the ionised
fraction $Q_{\delta,R}$ as a function of overdensity and
smoothing scale.  We repeatedly filter the linear density field at
logarithmic intervals on scales comparable to the box size down to the
grid scale size. For all filter scales, the ionisation state of each
grid position is determined using $Q_{\delta,R}$ and deemed to be
fully ionised if $Q_{\delta,R}\geqslant 1$. All voxels within a sphere
of radius $R$ centered on these positions are flagged and assigned
$Q_{\delta,R}=1$, while the remaining non-ionised voxels are assigned
an ionised fraction of $Q_{\delta,R_{\textrm{f,min}}}$, where
$R_{\textrm{f,min}}$ corresponds to the smallest smoothing scale. A
voxel forms part of an HII region if $Q_{\delta,R}>1$ on any scale
$R$. In this paper we present simulations corresponding to a linear
density field of resolution $256^{3}$, with a comoving side length of
512\,Mpc.  We label positions within the simulation with the vector
$\mathbf{x}=(x_1,x_2,x_3)$, and define the center of the box as the
origin. We take the $x_3$-axis to be parallel to the line of sight,
and show results using dimensions of proper distance.  The
line-of-sight coordinate ($x_3$) can be equivalently described using
redshift or frequency.  We have chosen the side length of
512\,comoving Mpc to match the 32\,MHz bandpass of the MWA, centered
on a quasar at $z=6.65$.

\subsection{The evolution of quasar HII regions in a percolating IGM}
\label{evolution}

A semi-numerical simulation generates a realisation of the
three-dimensional ionisation field at a particular value of proper
time \citepalias{geil07}. However, late in the reionisation era, where
we expect to observe the high-redshift quasars, the light travel time
across a quasar HII region may be comparable to the timescale over
which the IGM evolves significantly. This evolution may have a negative
impact on the observability of quasar HII regions. In this paper we
therefore describe semi-numerical calculations of an ionisation field
for two different cases of IGM evolution. In the first we assume a
non-evolving IGM with a slowly evolving quasar. In the second we
consider a model where the line-of-sight axis includes the finite light
travel time, and evolution of the IGM ionisation and quasar
luminosity. The first case corresponds to the model in
\citetalias{geil07}. The additions to this model that are required for
the second case (with an evolving IGM) are as follows.

Unlike an $N$-body simulation which evolves an ionisation field in time,
a semi-numerical calculation computes the ionisation field at a single
instant in time.  To include the finite speed of light we must
therefore compute many simulations at closely separated redshifts and
stack slices with line-of-sight distances such that the photons from
all slices reach the observer at the same time. The
redshift of a slice is
\begin{equation}
z \approx z_0+\left(\frac{cdt}{dz}\right)^{-1}(x_3-x_3^0),
\end{equation}
where $x_3$ is the proper distance coordinate along the line of sight,
and $x_{3}^0$ is proper the distance to redshift $z_0$.

\subsection{Inclusion of quasars in the semi-numerical scheme}
\label{quasar}

\begin{figure*}
\includegraphics[width=15.3cm]{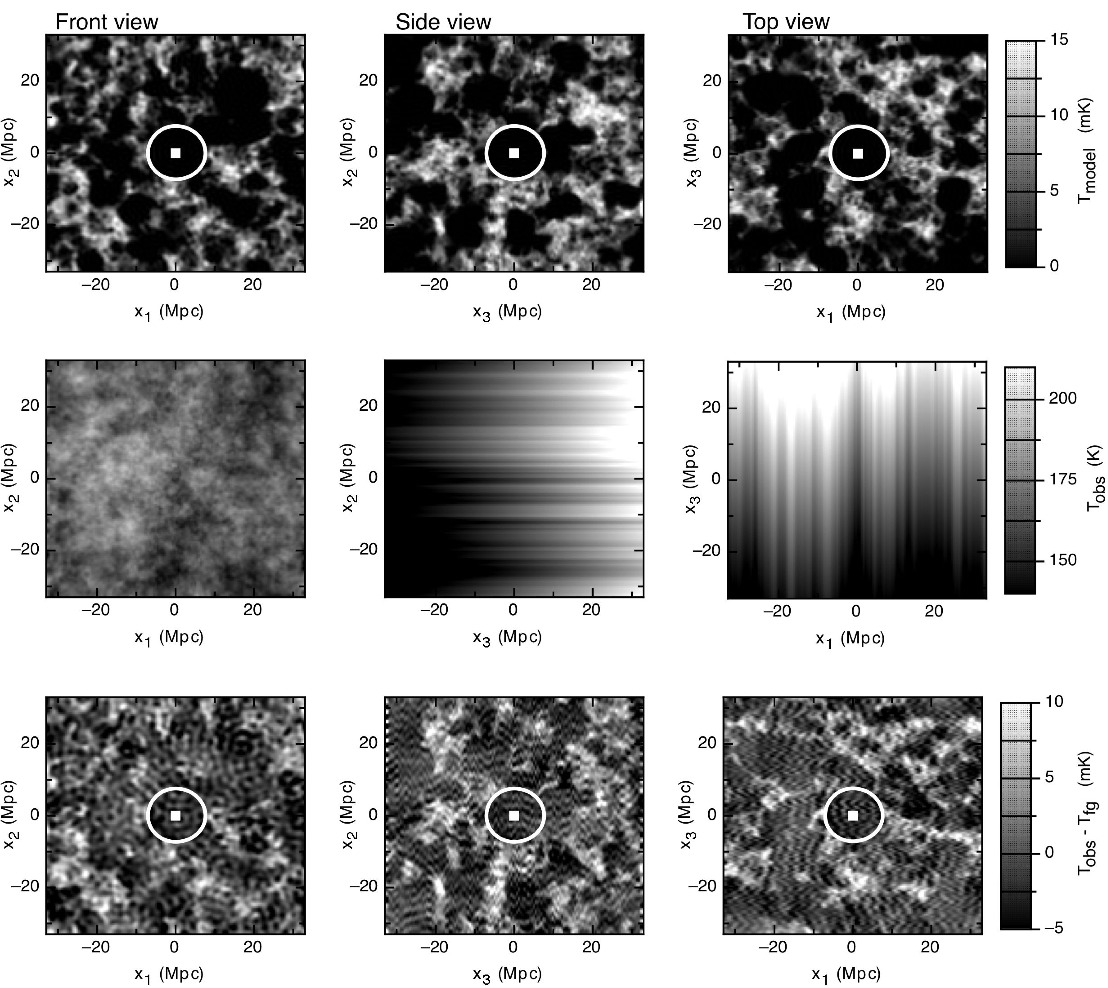} 
\caption{A quasar HII region in a non-evolving IGM. Three aspects of the HII region
are shown, with slices through the center of the box when viewed from the front, top
and side. Each slice is $6\,$Mpc thick, which corresponds to $\sim3\,$MHz along
the $x_3$-axis. In observed units, the cube is $\sim3.3$ degrees on a side and
$33\,$MHz deep. The model, foreground plus model, and observed maps following
foreground removal are shown in the upper, central and lower panels respectively.
The shape of the HII region extracted from the foreground removed cubes using the
method described in the text are also plotted. The quasar was assumed to have
$R_{\rm q,tot}=34$ comoving Mpc, and its position is plotted as the square point.
The mass-averaged IGM neutral fraction is assumed to be 15\%. For the purposes
of display we have used the central redshift of the box to convert between proper
and comoving distances throughout this paper.}
\label{fig1}
\end{figure*}

In this section we describe the effect of the quasar on the evolving ionisation
structure of the IGM.  The quasar has an influence which evolves with
time.  We assume the quasar turns on(off) at a redshift $z_{\rm
on(off)}$ (corresponding to a proper time $t_{\rm on(off)}$), and has
a lifetime $t_{\rm lt} = t_{\rm off} - t_{\rm on}$.  Following the
calculation in \citetalias{geil07} we parameterise the influence of
the quasar using the distance $R_{\rm q}$, which is the radius of the
region that would have been ionised by the quasar alone in a fully
neutral IGM
\begin{equation}
\label{quasartd}
R_{\rm q} = \left\{ \begin{array}{ll}
0 & z>z_{\rm on}\\
R_{\rm q,tot}\left(t/t_{\rm on}\right)^{1/3} & z_{\rm off}<z<z_{\rm on}\\
R_{\rm q,tot} & z<z_{\rm off},
\end{array} \right.
\end{equation}
where $R_{\rm q,tot}$ is the maximum radius of the HII region that
would have been generated by the quasar in a fully neutral IGM.

To include the influence of a quasar in an evolving IGM we generate an
ionisation cube for the IGM as a function of look-back time using the
procedure described above. However in doing so we use
equation~(\ref{quasartd}) to include the time-dependent quasar
contribution to the ionising photon budget.  In a region of radius $R$
containing a quasar, the cumulative number of ionisations per baryon
will be larger than predicted by the semi-analytic model. We include
quasars in our scheme by first computing the fraction of the IGM
within a region of radius $R$ centered on the position $\mathbf{x}$
that has already been ionised by stars.  We then add an additional
ionisation fraction equal to the quasar's contribution
\begin{equation} Q_{\rm q}(\mathbf{x}) =
\left(\frac{|\mathbf{x}-\mathbf{x}_{\rm q}|}{R_{\rm q}}\right)^{-3},
\end{equation} 
where $\mathbf{x}_{\rm q}$ is the position of the quasar, and $R_{\rm
q}$ is computed via equation~(\ref{quasartd}).  In contrast to stellar
ionisation the quasar contribution comes from a point source, and so
the contribution to $Q$ originates from a single voxel only (rather
than all voxels within $|\mathbf{x}-\mathbf{x}_{\rm q}|$). Following
this addition we filter the ionisation field as described in
\S\,\ref{ionisationfield}.

The value of $R_{\rm q,tot}$ is subject to large uncertainties,
including the quasar duty cycle, lifetime and luminosity. Assuming a
line-of-sight ionising photon emission rate of $\dot{N}_{\gamma}/4\pi$
photons per second per steradian, a uniform IGM and neglecting
recombinations, the comoving line-of-sight extent of a quasar's HII
region (from the quasar to the \textit{front} of the region) observed
at time \textit{t} is \citep{white2003}
\begin{equation}\label{RqWhite}
R_{\rm q,tot} \approx 30\,\textrm{Mpc}\;x_{\rm HI}^{-1/3}
\left(\frac{\dot{N}_{\gamma}}{10^{57}}\frac{t_{\rm lt}}{10^7 \textrm{yr}}\right)^{1/3}
\left(\frac{1+z}{7.5}\right)^{-1},
\end{equation}
where $x_{\rm HI}$ is the neutral fraction. In this paper we show
examples of $R_{\rm q,tot} \approx34$ comoving Mpc, consistent with
the lower limit around the luminous SDSS quasars at $z>6$
\citep{fan2006}.

\begin{figure*}
\includegraphics[width=15.3cm]{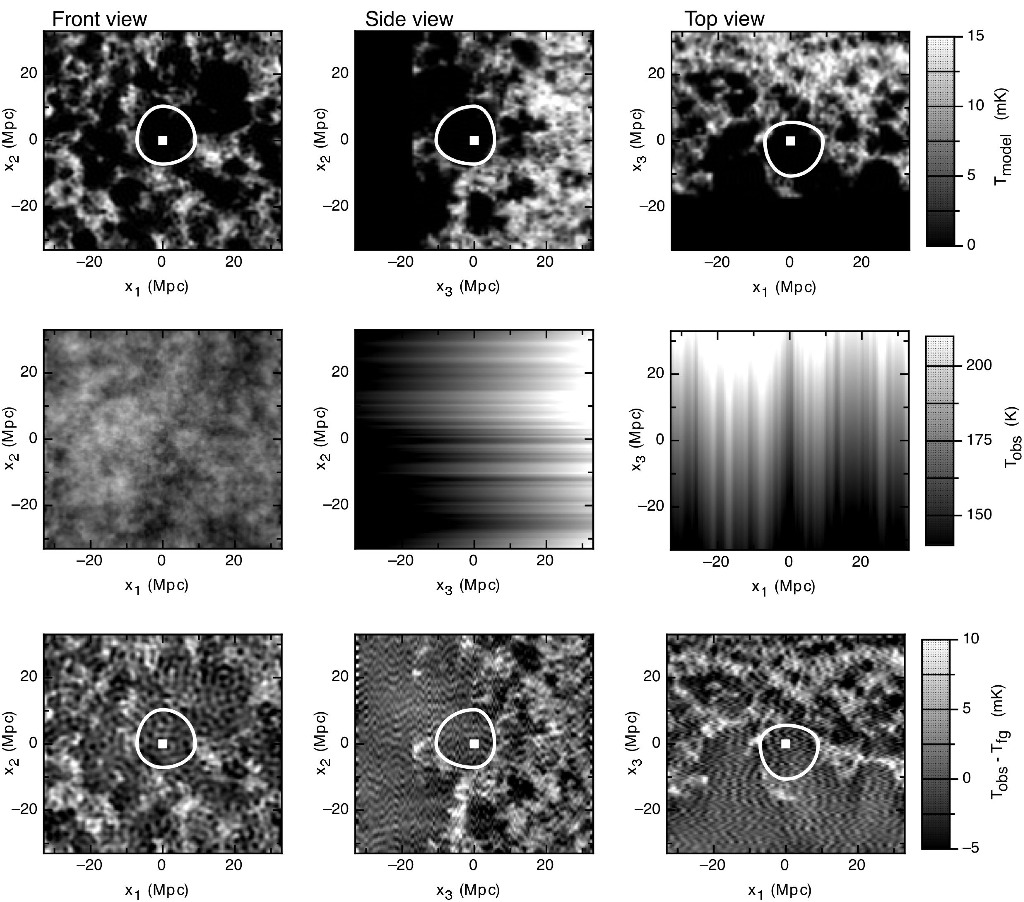} 
\caption{ As per Figure~\ref{fig1}, this time for a quasar in an evolving IGM.
The quasar was assumed to have $R_{\rm q,tot}=34$ comoving Mpc and a
lifetime of $4\times10^7$\,yr centered on $z=6.65$, which is also the redshift
at the center of the simulation box. The mass-averaged IGM neutral fraction
at the redshift of the quasar is $\approx15\%$.}
\label{fig2}
\end{figure*}

\section{Examples of quasar HII regions}
\label{results}

We compute the three-dimensional ionisation structure for our
reionisation model using the prescription outlined in \S\,\ref{model}.
The contrast in 21-cm brightness temperature corresponding to a region in
the simulation with ionisation fraction $Q$ and overdensity $\delta$ is
\begin{equation} T(\delta,R) =
22\mbox{ mK }(1-Q_{\delta,R})\left(\frac{1+z}{7.5}\right)^{1/2}\left(1+\delta\right).
\end{equation} 
Here we have assumed that the spin-temperature of neutral hydrogen in
the IGM is much larger than the cosmic microwave background (CMB)
temperature \citep{ciardi2003}, and ignored the enhancement of the
brightness temperature fluctuations due to peculiar velocities in
overdense regions \citep{bharadwaj2005,bl2005}. Because the
ionisation state of the IGM around the highest redshift quasars is not
known, we consider the two cases described in \S\,\ref{evolution}
which bracket the ionisation conditions surrounding a quasar prior to
the completion of reionisation.
  
\begin{figure*}
\includegraphics[width=17.0cm]{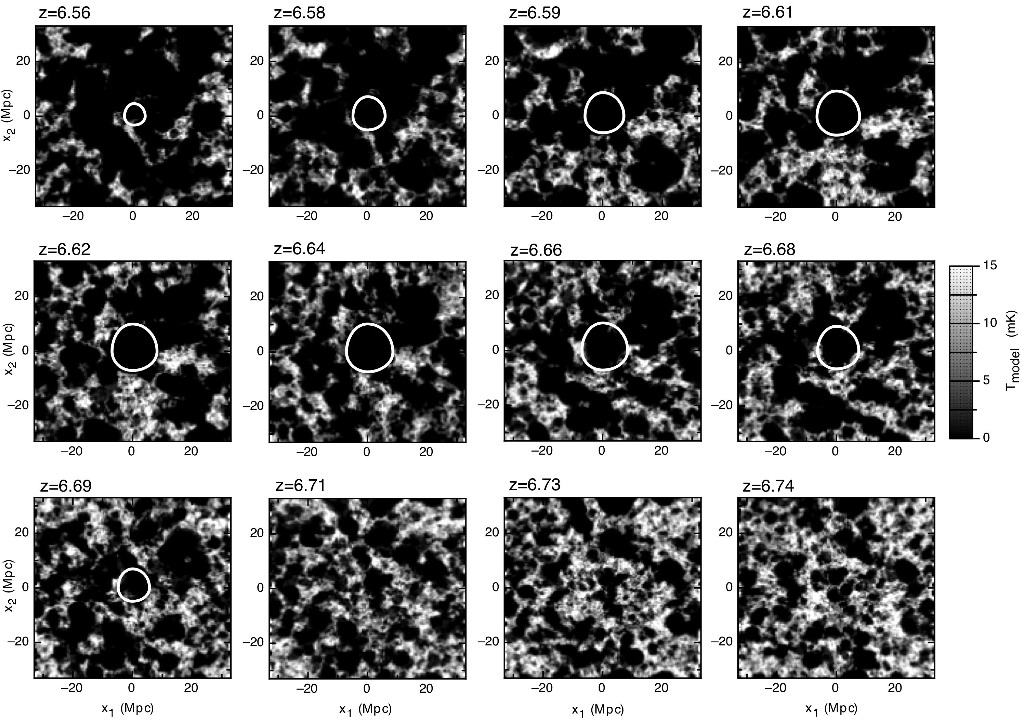} 
\caption{ Slices through the simulation of a quasar in an evolving IGM for
$6.56 \le z \le 6.74$, viewed from the front. The shapes of the HII regions
extracted from foreground removed cubes using the method described
in the text are also plotted.}
\label{fig3}
\end{figure*}

Resulting 21-cm brightness temperature maps are presented in
Figures~\ref{fig1}--\ref{fig3}.  In the upper row of Figure~\ref{fig1}
we show snapshots of the model brightness temperature $T_{\rm model}$
for slices through a realisation at a time when the mean neutral
fraction of the IGM is $\approx15\%$, corresponding to redshift
$z=6.65$ (assuming $R_{\rm q}=R_{\rm q,tot}$). This simulation was
computed at fixed proper time, and so represents the model with slow
evolution in the reionisation process. Each slice is 3 comoving Mpc
deep. We show three aspects of the HII region, with slices viewed from
the front, side and top. These slices are similar
because the simulation is statistically isotropic. The position of the
quasar within the simulation cube is plotted as a square point.

In Figures~\ref{fig2} and \ref{fig3} we show results for the model in
which finite light-travel time is included, showing the effect of
rapid evolution in the ionisation state of the IGM towards the end of the
reionisation era. In this simulation the center of the box is at $z=6.65$,
at which time the IGM has a neutral fraction of $\approx15\%$.  The
quasar is assumed to have $R_{\rm q,tot}=34$ comoving Mpc, a
lifetime of $2\times 10^7$\,yr and be centered on $z=6.65$.  The position of
the quasar within the simulation cube is plotted as a square point.
The upper row of Figure~\ref{fig2} shows three aspects of the HII
region in this model, with slices viewed from the front, side
and top. The evolution of the IGM is clearly seen in this
figure, with the percolation process completing between the ``back" of
the box ($x_3>0$) and the ``front" of the box ($x_3<0$).  In
Figure~\ref{fig3} we show a sequence of slices between $z=6.74$ and
6.56. These slices show the thinning of neutral gas towards
lower redshifts. The presence of an HII region around the quasar can be
seen towards the middle of the redshift range.

\section{The Continuum Foreground}
\label{foreground}

Foreground contamination and its removal could have significant
consequences for the detectability of the weak, redshifted 21-cm
signal. Indeed foreground contamination will be brighter than the
cosmological 21-cm signal by 4--5 orders of magnitude. The three main
sources of foreground contamination of the 21-cm signal are Galactic
synchrotron (which comprises $\sim$\,70$\%$), extragalactic point
sources (27$\%$) and Galactic bremsstrahlung (1$\%$)
\citep{shaver1999}. The frequency dependence of these foregrounds can
be approximated by power laws with a running spectral
index \citep{shaver1999,tegmark2000}. While the sum of power laws is
not in general a power law, over a relatively narrow frequency range
(such as that considered in this paper where $\Delta\nu/\nu\ll 1$), a
Taylor expansion around a power law can be used to describe the spectral shape.
We therefore also approximate the sum of foregrounds as a power law with a running
spectral index, and specialise to the case of Galactic synchrotron
emission, which dominates the foregrounds\footnote{The primary effect
of other foregrounds will be to slightly alter the spectral index and
the frequency dependence of the spectral index, and also (in the case
of unresolved radio point sources) the angular structure of the
foreground at small scales. Neither of these effects is important as
our foreground removal technique is not sensitive to the specific
value of the spectral index we adopt. As for angular fluctuations,
these correspond to zero-point fluctuations along different lines of
sight, which are immediately removed by foreground cleaning. We have
experimented with increasing angular fluctuations by a factor of 10,
and found little difference.}. For more detailed foreground models,
including synchrotron emission from discrete sources such as supernova
remnants, and free-free emission from diffuse ionised gas, see
\cite{jelic08}. We assume that the brightest point sources have been
removed and only consider unpolarised foregrounds.

The intensity of Galactic synchrotron emission varies as a
function of both sky position and frequency. We model the frequency and
angular dependence of Galactic synchrotron foreground emission as follows.
We first construct a realisation of the angular fluctuations in the foreground
(around the mean brightness temperature) at a particular frequency $\nu_0$ using the relation
\begin{equation}
\frac{l^2 C_l(\nu_0)}{2\pi}=\left(\frac{l}{l_0}\right)^{2-\beta} T_{l_0}^{\rm syn}(\nu_0)^{2},
\end{equation}
where the frequency dependence of the fluctuation amplitude is given by
\begin{equation}
T_{l_{0}}^{\rm syn}(\nu_0)=A_{l_{0}}^{\rm syn} \bigg( \frac{\nu_0}{\rm 150\,MHz}\bigg)
^{-\alpha_{\rm syn}-\Delta \alpha_{\rm syn} \rm log_{10}[\nu_0/(\rm 150\,MHz)]},
\end{equation}
and the constants have values $\alpha_{\rm syn}=2.55$, $\Delta \alpha_{\rm syn}=0.1$ and $A_{l_{0}}^{\rm
syn}=25$\,K \citep{shaver1999,tegmark2000,wang2006}. The latter two
values are extrapolated from 30 GHz CMB observations
\citep{mcquinn2006} since at present there are no data at lower
frequencies with arc minute resolution\footnote{Although the high frequencies of
CMB experiments make them less than ideal for understanding diffuse Galactic
emission at frequencies relevent to high-redshift 21-cm tomography, several
publicly available unpolarised, large area radio surveys have been compiled
to form a Global Sky Model (see e.g. http://space.mit.edu/$\sim$angelica/gsm/index.html).
This, together with the 843\,MHz sub-degree resolution and $\sim1$\,Jy\,beam$^{-1}$
rms sensitivity SUMMS survey \citep{bock1999} for $\delta<-30^{\circ}$, will provide a
better practical continuum foreground model for the MWA target fields
(Randall Wayth, private communication).}. We then add a mean sky brightness
$\bar{T}^{\rm syn}=165\,(\nu_0/\mbox{185\,MHz})^{-2.6}$\,K to our two-dimensional
realisation of brightness temperature fluctuations $\Delta T^{\rm syn}(\mathbf{\theta})$
(although interferometers will generally not be sensitive to the temperature
zero-point). This foreground plane is then extended into three dimensions by
extrapolation, using the running power law form along each line of sight
\begin{equation}
T^{\rm syn}(\mathbf{\theta},\nu) = \left[\bar{T}^{\rm syn} + \Delta T^{\rm syn}(\mathbf{\theta})
\right] \bigg( \frac{\nu}{\nu_0}\bigg)^{-\alpha_{\rm syn}-\Delta \alpha_{\rm syn} \rm log_{10}[\nu/\nu_0]}.
\end{equation}
In this paper we have not modeled the spatial variation of $\alpha$
and $\Delta \alpha_{\rm syn}$, although the effect of this is
negligible, as we have verified in other work (Petrovic \& Oh, 2008,
in preparation). Most of the error in foreground fitting comes from
``noise" due to the instrument and the cosmological 21-cm signal.

The total observed brightness temperature becomes $T_{\rm
obs}=T^{\rm syn}+T_{\rm model}$. Maps of the continuum
foreground model plus 21-cm signal are shown in the second row of panels
in Figures~\ref{fig1} and \ref{fig2}. 
The foreground completely swamps the HII region
signal, which is not visible in the combined maps. 
The mean and variance of the foreground brightness (at the central frequency
of these cubes) are $1.7\times10^5$\,mK and $8.7\times10^3$\,mK respectively,
which should be compared to the  $\sim10$\,mK 21-cm fluctuations. The front,
side and top views are shown as before. The front view shows the angular
structure of the foreground, while the side and top views show
the coherent structure along the line of sight, owing to the smooth
frequency dependence of the assumed foreground emission.

\section{Synthetic Observations of Model HII Regions}
\label{obs}

In this section we discuss the response of a low-frequency array to
the model HII regions described in \S\,\ref{results}, including
foreground subtraction. We then estimate the potential for measurement
of HII region properties, including the HII region shape and the
global neutral fraction.

\subsection{Fundamental sensitivity limit to the 21-cm signal}
\label{sensitivity}

To compute the response of a low-frequency array to the model HII
region we follow the prescription given in \citetalias{geil07}.  Radio
interferometers measure a frequency-dependent, complex visibility
$V(\mathbf{U},\nu)$ (Jy) for each frequency channel and baseline
$\mathbf{U}$ in their configuration. The measured visibility is, in
general, a linear combination of signal, foreground and system noise,
\begin{equation}
\label{vis}
V(\mathbf{U},\nu) = V_{\rm S}(\mathbf{U},\nu) + V_{\rm F}(\mathbf{U},\nu) + V_{\rm N}(\mathbf{U},\nu),
\end{equation}
where $V_{\rm S}$ is the signal, $V_{\rm F}$ the contribution due to
foreground sources and $V_{\rm N}$ the system noise. The root mean
square of noise fluctuation per visibility per frequency channel is
given by
\begin{equation}
\label{radiometer}
\Delta V_{\rm N} = \frac{2k_{\rm B}T_{\rm sys}}{A_{\rm eff}\sqrt{t_{\mathbf{U}}\Delta\nu}},
\end{equation}
where $T_{\rm sys}$ is the system temperature (K), $A_{\rm eff}$ the
effective area of one antenna (m$^{2}$), $t_{\mathbf{U}}$ is the
integration time for that visibility (s), $\Delta\nu$ is the frequency
bin width (Hz) and $k_{\rm B}$ is the Boltzmann constant. Instrumental
noise is uncorrelated in the frequency domain and is assumed to be
Gaussian. As shown by \cite{mcquinn2006}, the average integration time
$t_{\mathbf{U}}$ that an array observes the visibility $\mathbf{U}$ is
\begin{equation}
\label{tU}
t_{\mathbf{U}} \approx \frac{A_{\rm eff}t_{\rm int}}{\lambda^{2}}n(\mathbf{U}),
\end{equation}
where $t_{\rm int}$ is the total integration time and $n(\mathbf{U})$
is the number density of baselines that can observe the visibility
$\mathbf{U}$. The transverse wavenumber $k_{\perp}$ is given by
$k_{\perp} = 2\pi |\mathbf{U}|/r_{\rm em}(z)$, where $r_{\rm em}(z)$
is the proper distance to the point of emission.

We present simulated results for the MWA, which will consist of 512
tiles, each with 16 cross-dipoles with 1.07\,m spacing and an
effective area $A_{\rm eff}\approx16(\lambda^{2}/4)$ for
$\lambda\lsim$ 2.1\,m. The system temperature at $\nu<200$\,MHz is sky
dominated and has a value $T_{\rm
sys}\sim250\left[(1+z)/7\right]^{2.6}$\,K. This value is an estimate
of the average sky brightness due to Galactic synchrotron emission
over the primary beam. Following \cite{bowman2006}, we assume a smooth
antenna density profile $\rho_{\rm{ant}}\propto r^{-2}$ within a 750 m
radius. For the MWA, the aperture is filled within a radius of
18\,m. In order to maintain high sensitivity we truncate the naturally
weighted visibility data using a filled circular aperture of radius
$a$ with a corresponding beamwidth (full width at half maximum)
$\theta_{\rm b} = 0.705U_{\rm max}^{-1}$, where $U_{\rm max} =
a/\lambda$. In this paper we show results computed for $\theta_{\rm b}
= 3.2'$, and assume an integration time of $t_{\rm int}=100$ hours.

We simulate the thermal noise in a three-dimensional
visibility-frequency cube using equations (\ref{radiometer}) and
(\ref{tU}), allowing us to account for a particular array
configuration and observation strategy. We then perform a
two-dimensional inverse Fourier transform in the $uv$-plane for each
frequency channel in the bandwidth, which gives a realisation of the
system noise in the image cube (i.e. sky coordinates). Using the
linearity of both the Fourier transform and equation (\ref{vis}), the
observed specific intensity can then be found either by first adding
visibility terms and then transforming to image space, or by adding
the realisation of each visibility component in image space. Finally,
we apply the Rayleigh-Jeans approximation to express the observed flux
in terms of brightness temperature $T$ using $\partial
B_{\nu}/\partial T = 2k_{\rm B}/\lambda^{2}$, where $B_{\nu}$ is the
specific intensity.

\subsection{Continuum foreground subtraction}
\label{foregrounds}

It has been suggested that spectrally smooth foregrounds could be
removed through continuum subtraction
\citep{gnedin2004,wang2006,jelic08}.  We model the continuum using
a polynomial of the form
\begin{equation}
\log_{10}(T_{\rm fg})=\sum_{i=0}^{n_{\rm log}} C_i [\log_{10}(\nu)]^i,
\end{equation}
where $\nu$ is the observed frequency\footnote{Note that when compared
with the frequency evolution assumed for our synchrotron model, the
coefficients $C_1$ and $C_2$ correspond to $\alpha_{\rm syn}$ and
$\Delta\alpha_{\rm syn}$ respectively.}.  We begin with results for
$n_{\rm log}=3$ (which we label model II), but return to discuss
different foreground removal models in \S~\ref{fg}.

We perform a fit of this functional form along the line of sight to
each spatial pixel in the simulation cube. We then subtract the
best-fit, leaving residual fluctuations around the foreground
emission. The resulting foreground subtracted maps are plotted in the
lower rows of Figures~\ref{fig1} and \ref{fig2}. Inspection of these
maps indicates that the main features of the HII regions are still
present. However, foreground removal has significantly lowered the
contrast of the maps relative to the input model.  This is because
foreground removal erases modes in the signal with wavelengths
comparable to the bandpass, leaving a foreground-cleaned data cube
composed of residuals, with a near zero-mean brightness temperature.

\subsection{The shape of HII regions}

\begin{table}
\begin{center}
\caption{\label{tab1} Parameters describing the HII region shape in models
of non-evolving and evolving IGM, corresponding to Figures~\ref{fig1} and
\ref{fig2} respectively. The values in parentheses are the offset relative to
the volume-averaged radius $\langle R^3\rangle^{1/3}$.}
\begin{tabular}{ccccc}
\hline
&\multicolumn{2}{c}{evolving IGM}&\multicolumn{2}{c}{non-evolving IGM}\\
&FGR map&model&FGR map&model\\\hline
$A_1$&7.1 (-1.7)&7.0 (-1.6)&7.1 (-0.2)&6.8 (-0.6)\\
$A_2$&9.0 (+0.2)&8.4 (-0.2)&7.6 (+0.1)&7.6 (+0.2)\\
$B_1$&7.1 (-1.7)&7.4 (-1.2)&7.3 (-0.1)&7.3 (-0.1)\\
$B_2$&10.3 (+1.5)&10.4 (+1.8)&7.6 (+0.2)&7.5 (+0.1)\\
$C_1$&11.0 (+2.2)&10.5 (+1.9)&7.1 (-0.3)&7.4 (0.0)\\
$C_2$&5.6 (-3.2)&5.6 (-3.0)&7.6 (+0.5)&7.6 (+0.2)\\
$\langle R^3\rangle^{1/3}$&8.8&8.6&7.4&7.4 \\\hline
\end{tabular}
\end{center}
\end{table}

In our model we have assumed isotropic quasar emission, which
is probably not correct. However even in this case the resulting HII
region may not be spherical owing to the inhomogeneous nature of the
stellar reionisation around the quasar HII region, as well as the
effects of evolution which could serve to elongate the HII region
along the line of sight \citep{wlb2005,yu05}.

In an inhomogeneously ionised IGM the meaning of ``shape" is not
clear. For the purpose of illustration, we define the shape of the HII
region to be the shape of a spheroid that maximises the contrast in 21-cm emission,
\begin{equation}
\label{contrast}
T_{\rm contrast} = \langle T_{\rm out}\rangle - \langle T_{\rm in}\rangle,  
\end{equation}
between two concentric shells, where $\langle T_{\rm out}\rangle$ and
$\langle T_{\rm in}\rangle$ are the mean emission within shells of
thickness $\Delta R$ outside and inside a spheroidal surface
respectively.  We then parameterise this spheroid by
\begin{equation}
\label{sph}
\bigg(\frac{x_1}{a}\bigg)^2 + \bigg(\frac{x_2}{b} \bigg)^2 + \bigg(\frac{x_3}{c} \bigg)^2 = 1,
\end{equation}
where $a=a_1$ for $x_1<0$ and $a=a_2$ for $x_1>0$, $b=b_1$ for $x_2<0$
and $b=b_2$ for $x_2>0$, and $c=c_1$ for $x_3<0$ and $c=c_2$ for
$x_3>0$. To estimate the shape of the HII region we therefore maximise
the quantity $T_{\rm contrast}$ by varying the parameter set
$\mathbf{p} = (a_1,a_2,b_1,b_2,c_1,c_2)$. We assume $\Delta R=1\,$Mpc
and that the spheroid is centered on the quasar (we also assume the position
and redshift of the quasar has been accurately measured by optical/IR
observations). In order to evaluate the prospect of measuring the shape
of a quasar HII region, this maximisation is performed in the noisy, foreground
subtracted data sets shown in Figures~\ref{fig1} and \ref{fig2}.

The resulting optimised spheroids are plotted over both model and
foreground subtracted noisy maps in
Figures~\ref{fig1} and \ref{fig2}. The numerical values of the fitted
parameters $\mathbf{p}$ are listed in Table~\ref{tab1}. Although the
maximisations were performed in the foreground removed maps, the
shapes of the recovered HII regions faithfully represent the input
model. For example, in the evolving IGM model (Figure~\ref{fig2})
the fitted spheroid describes the extension of the observed HII region
towards the observer along the (negative) $x_3$-direction. This
faithful reproduction holds for both the non-evolving and evolving
models, and indicates that foreground removal does not erase
information regarding HII region shape. To quantify this statement we
also perform the maximisation of $T_{\rm contrast}$ on the
corresponding input models and list the resulting parameters
$\mathbf{p}$ in Table~\ref{tab1} for comparison. The shape of the HII
region, including the asymmetries, are very similar in the two
cases. We also calculate the volume-averaged radii $\langle
R^3\rangle^{1/3}$ and list these values in Table~\ref{tab1}. The
values in parentheses are the offset of the parameters $\mathbf{p}$
with respect to this volume-averaged radius. These offsets are similar between
the input model and noisy foreground maps, reaffirming that the
foreground removal process does not degrade the ability to measure HII
region shape and size.

\begin{table}
\begin{center}
\caption{\label{tab2} Values of recovered neutral fraction for the four foreground
models. The model values are listed in parentheses.}
\begin{tabular}{cccc}
\hline
model& index     & evolving IGM   & non-evolving IGM  \\\hline
I & $n_{\rm log}=2$ & 0.07 (0.15) & 0.10 (0.14)\\
II & $n_{\rm log}=3$ & 0.07 (0.15) & 0.10 (0.14)\\
III & $n_{\rm log}=4$ & 0.04 (0.15) & 0.04 (0.14)\\
IV  &  $n_{\rm lin}=4$ & 0.04 (0.15) & 0.05 (0.14)\\\hline
\end{tabular}
\end{center}
\end{table}

\subsection{Estimate of neutral fraction}
\label{neutfrac}

Based on the fact that the HII region shape is not known a priori,
\citetalias{geil07} investigated the constraints on neutral fraction
that could be placed using a narrow beam along the quasar line of
sight. However, in the previous section we have argued that the
spheroidal shape of the HII region can be accurately reproduced from
foreground subtracted maps. We can therefore use the full data cube to
determine the contrast of emission that originates from regions that
are inside and outside of the HII region respectively, rather than
just along one line of sight. To achieve this we fit planes of 21-cm emission 
\begin{equation}
\label{Tin}
T_{\rm in}(\mathbf{x}) = \bar{T}_{\rm in} + A_{\rm in}\,x_1 + B_{\rm in}\,x_2 + C_{\rm in}\,x_3
\end{equation}
and 
\begin{equation}
\label{Tout}
T_{\rm out}(\mathbf{x}) = \bar{T}_{\rm out} + A_{\rm out}\,x_1 + B_{\rm out}\,x_2 + C_{\rm out}\,x_3
\end{equation}
to regions of the simulations of foreground subtracted 21-cm intensity
$T(\mathbf{x}) = T_{\rm obs}(\mathbf{x})-T_{\rm fg}(\mathbf{x})$ that
are inside and outside the HII region respectively.
This method was adopted in order to fit for both the mean value and evolution of
global neutral fraction. Using these solutions we can estimate the global mass-averaged
neutral fraction at the redshift of the quasar, which is obtained from
\begin{equation}
x_{\rm HI} = \frac{\bar{T}_{\rm out}-\bar{T}_{\rm in}}{22\mbox{mK}}\left(\frac{1+z}{7.5}\right)^{-1/2}.
\end{equation}
Performing these fits for the models shown in Figures~\ref{fig1} and
\ref{fig2} gives $x_{\rm HI}= 0.10$ and 0.07 respectively (see
Table~\ref{tab2}).
Ideally, the coefficients $A_{\rm in/out}$ and $B_{\rm in/out}$ should vanish.
The value of $\Delta C = C_{\rm out}-C_{\rm out}$ could in principle be used to measure the
evolution in global neutral fraction, however noise and foreground removal will
prevent this measurement in practice.

For comparison, we have also computed the
corresponding neutral fractions from fits to the input model, which
yield values of $x_{\rm HI}= 0.14$ and 0.15 respectively. Thus,
foreground removal erases a significant fraction of the contrast
between the HII region and the IGM, which quantifies the observations
made from inspection of Figures~\ref{fig1} and \ref{fig2} in earlier
sections. We note that the fraction of contrast that is lost
differs between these two examples, owing to the large difference in
evolutionary properties of the IGM. This suggests that the effect of
foreground removal on the inferred neutral fraction could only be
reliably estimated via detailed numerical modelling.

\section{Sensitivity to the foreground removal model}
\label{fg}

Before concluding, we investigate the effect of different foreground
removal models on the HII region observables. Specifically, in
addition to model II ($n_{\rm log}=3$), we choose polynomials in
$\log_{10}(\nu)$ with $n_{\rm log}=2$ (labelled model I, corresponding
to the form of the assumed synchrotron foreground) and $n_{\rm log}=4$
(labelled model III). We also consider a polynomial in $\nu$
[as opposed to $\log_{10}(\nu)$] of degree $n_{\rm lin}=4$ (labelled
model IV).

We find that results for the parameters $\mathbf{p}$ describing the
shape of the HII region are insensitive to the foreground model, and
so have not listed additional results for models I, III and
IV. However we find that the neutral fraction derived from
observation of an HII region is sensitive to the foreground model
assumed, and list these results in Table~\ref{tab2}. Models I and
II (with $n_{\rm log}=2$ and 3 respectively) yield the same neutral
fraction ($x_{\rm HI}=0.07$ and $0.10$ for evolving and non-evolving
models respectively). However, the models of higher order (models III and
IV) remove additional contrast from the HII region, leading to
smaller recovered neutral fractions. This indicates that a model which provides for
fluctuations that are of a higher order than those present
in the true foreground will lead to removal of power from the inferred 21-cm
signal. This in turn implies that the foregrounds will need to be very
well understood in order to make quantitative measurements of the
neutral fraction based on the observation of HII regions.

\section{Conclusion}
\label{conclusion}

One of the primary goals of future low-frequency telescopes is the
detection of large quasar-generated HII regions in redshifted 21-cm
emission during the epoch of reionisation. Like all 21-cm signatures of the
reionisation era, the detection of a quasar HII region will need to
overcome the difficulties associated with the removal of bright
Galactic and extragalactic foregrounds. In this paper we have assessed
the impact of removing a spectrally smooth foreground on the
observable properties of high-redshift quasar HII regions. We have
assumed prior removal of extragalactic point sources and considered
only the unpolarised Galactic synchrotron foreground.

The primary effect of continuum foreground removal is to erase
contrast in the image. In particular, contributions to the 21-cm
intensity fluctuations that have a scale length comparable to the
frequency bandpass of the observation are removed by foreground
subtraction.  On the other hand, we find that this loss of contrast
does not affect the ability of 21-cm observations to measure the size
and shape of the HII region. We model the HII region shape as a
spheroid described by six parameters and show that the shape recovered
following foreground removal agrees well with the shape derived
directly from fitting using the input model.

Using the recovered best-fit shape of the HII region, the global neutral fraction
of hydrogen in the IGM could be measured directly from the contrast in intensity
between regions that are within and beyond the HII region. However, we find
that since foreground removal lowers the observed contrast between the
HII region and the IGM, such a measurement of the neutral fraction
would require a correction factor. Our results suggest that the value
of this correction factor depends on the reionisation history. This
correction factor would therefore need to be modeled using uncertain
astrophysics. In addition, the measured contrast of the HII region, and
therefore the inferred neutral fraction, is sensitive to the degree of the
polynomial used for the foreground removal. Thus, measurement of the
neutral fraction from quasar HII regions will require a detailed
knowledge of the continuum foreground spectrum.

Finally, we have considered both cases where the quasar is
observed in an IGM which evolves slowly relative to the light crossing
time of the HII region, and where the IGM evolves rapidly. The latter
case is likely to be more relevant for observations around
high-redshift quasars which are observed to be very rare at $z>6$. Our
results indicate that the evolution of the IGM will not impact
negatively on the ability of 21-cm observations to measure the size
and shape of quasar HII regions.

\bigskip

{\bf Acknowledgments} PMG acknowledges the support of an Australian Postgraduate
Award. The research was supported by the Australian Research Council (JSBW). NP
and SPO acknowledge support from NSF grant AST-0407084 and NASA grant NNG06H95G.

\newcommand{\noopsort}[1]{}

\bibliography{bib}

\label{lastpage}
\end{document}